\documentclass[aps,prl,twocolumn,superscriptaddress,showpacs]{revtex4}
\usepackage{graphicx}

\usepackage{dcolumn}

\usepackage{bm}
\usepackage{color}
\usepackage[colorlinks]{hyperref}
\input{epsf}
\begin{document}

%\preprint{}

\title{Phase study of oscillatory resistances in high mobility GaAs/AlGaAs devices:
Indications of a new class of integral quantum Hall effect}

\author{R. G. Mani}
\affiliation{Harvard University, Gordon McKay Laboratory of
Applied Science, 9 Oxford Street, Cambridge, MA 02138 U.S.A.}

\author{W. B. Johnson}
 \affiliation{Laboratory for Physical Sciences, University of
 Maryland, College Park, MD 20740 U.S.A.}

\author{V. Umansky}
\affiliation{Braun Center for Submicron Research, Weizmann
Institute, Rehovot 76100, Israel}

\author{V. Narayanamurti}
\affiliation{Harvard University, Gordon McKay Laboratory of
Applied Science, 9 Oxford Street, Cambridge, MA 02138 U.S.A.}

\author{K. Ploog}
\affiliation{Paul-Drude-Institut f\"{u}r Festk\"{o}rperelektronik,
Hausvogteiplatz 5-7, 10117 Berlin, Germany}

%\author{}
%\thanks{}
%\affiliation{}
\date{\today}
\begin{abstract}
An experimental study of the high mobility GaAs/AlGaAs system at
large-$\nu$ indicates several distinct phase relations between the
oscillatory diagonal- and Hall- resistances, and suggests a new
class of integral quantum Hall effect, which is characterized by
"anti-phase" Hall- and diagonal- resistance oscillations.
\end{abstract}
%\pacs{}
%\keywords{}
\maketitle A 2-dimensional electron system (2DES) at high magnetic
fields, $B$, and low temperatures, $T$, exhibits the integral
quantum Hall effect (IQHE), which is characterized by plateaus in
the Hall resistance $R_{xy}$ vs. $B$, at $R_{xy} = h/ie^{2}$, with
$i$ = 1,2,3,... and concurrent vanishing diagonal resistance
$R_{xx}$ as $T \rightarrow$ 0 K, in the vicinity of integral
filling factors of Landau levels, i.e., $\nu \approx
i$.\cite{1,2,3} With the increase of the electron mobility, $\mu$,
at a given electron density, $n$, and $T$, IQHE plateaus typically
become narrower as fractional quantum Hall effects (FQHE) appear
in the vicinity of $\nu \approx p/q$, at $R_{xy} =
h/[(p/q)e^{2}]$, where $p/q$ denotes mostly odd-denominator
rational fractions.\cite{2,3} Experimental studies of the highest
mobility specimens have mostly focused upon FQHE and other novel
phases.\cite{2,3,4,5} Meanwhile, the possibility of new variations
of IQHE that might appear with the canonical effect in the
reduced-disorder specimen, especially at large-$\nu$, has been
largely unanticipated by experiment. Here, we show that three
distinct phase relationships can occur between the oscillatory
diagonal- and Hall- resistances in the high-mobility specimen at
$\nu > 5$, and that IQHE can be manifested in two of these
variations. The results therefore identify one new class of IQHE,
as they provide insight into the origin of oscillatory variations
in the Hall effect, and their evolution into plateaus, in the
low-$B$ large-$\nu$ regime of the radiation-induced
zero-resistance states in the \textit{photoexcited} high mobility
2DES.\cite{6,7,8} Thus, these experiments also serve to
characterize, classify, and clarify the large-$\nu$ dark
properties. The results recall previous suggestions of new
phenomena in the $\nu >> 1$ limit.\cite{9}
\begin{figure}[b]
%h=here, t=top, b=bottom, p=separate figure page
\begin{center}
\leavevmode \epsfxsize=3.25in
 \epsfbox {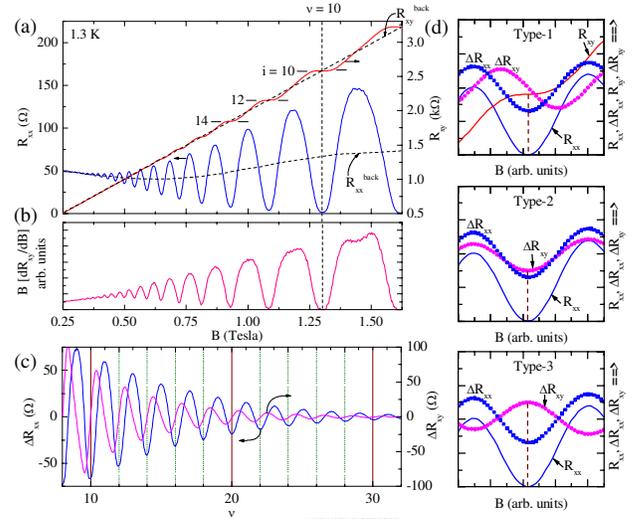}
\end{center}
\caption{ (a) Integral quantum Hall plateaus, at $R_{xy} =
h/ie^{2}$, coincide with resistance minima in $R_{xx}$ in a
GaAs/AlGaAs Hall bar device. (b) A comparison of  $B[dR_{xy}/dB]$
shown here with $R_{xx}$ in Fig. 1(a) suggests that $R_{xx}
\propto B[dRxy/dB]$, as per the resistivity rule (ref. 10). (c) A
quantum Hall system at large-$\nu$ also exhibits an approximately
$90^{0}$ phase shift between the oscillatory parts of $R_{xx}$ and
$R_{xy}$. (d) This panel illustrates the three types of
magnetoresistance oscillations that are reported in this study.}
\label{mani01fig}
\end{figure}

Figure 1(a) exhibits data from a low mobility Hall bar specimen
with $n = 3.2 \times 10^{11} cm^{-2}$ and $\mu = 0.4 \times 10^{6}
cm^{2}/V-s$. Here, large amplitude Shubnikov-de Haas (SdH)
oscillations in $R_{xx}$ lead into zero-resistance states with
increasing $B$, as $R_{xy}$ exhibits plateaus at $R_{xy} =
h/ie^{2}$ for $\nu \approx i$, with $i$ = 2,4,6,... This canonical
low-mobility IQHE system is known to exhibit a
resistivity/resistance rule,\cite{10,11} at a each $T$,\cite{12}
whereby $R_{xx} \propto B[dR_{xy}/dB]$ and $dR_{xy}/dB$ is the
$B$-field derivative of $R_{xy}$.\cite{10} Indeed, a comparison of
$R_{xx}$ (Fig. 1(a)) and $B[dR_{xy} /dB]$ (Fig. 1(b)) confirms
agreement.\cite{10,11,12,13}

For the sake of further analysis, Fig. 1(c) shows the oscillatory
part of the diagonal ($\Delta R_{xx}$) and Hall ($\Delta R_{xy}$)
resistances vs. $\nu$. Here, $\Delta R_{xy} = R_{xy} -
R_{xy}^{back}$ and $\Delta R_{xx} = R_{xx} - R_{xx}^{back}$, as
$R_{xy}^{back}$ and $R_{xx}^{back}$ are the background resistances
shown in Fig. 1(a). As evident from Fig. 1(c), the quantum Hall
characteristics of Fig. 1(a) yield approximately orthogonal
oscillations in $\Delta R_{xx}$ and $\Delta R_{xy}$ such that
$\Delta R_{xx} \approx -cos(2\pi[\nu/2])$ and $\Delta R_{xy}
\approx sin(2\pi[\nu/2])$. We denote the quantum Hall
characteristics of Fig. 1(a)-(c) as "Type-1" characteristics, and
sketch the essentials at the top of Fig. 1(d).

This study reports on other observable phase relations in the high
mobility 2DES. We find, for instance, a "Type-2" case, where
$|R_{xy}|$ is enhanced at the $R_{xx}$ oscillation peaks and the
$\Delta R_{xy}$ oscillations are in-phase with the $R_{xx}$ or
$\Delta R_{xx}$ oscillations, as illustrated in Fig. 1(d)
[center]. There also occurs the "Type-3" case where $|R_{xy}|$ is
suppressed at the $R_{xx}$ oscillation maxima and the $\Delta
R_{xy}$ oscillations are phase-shifted by "$\pi$" with respect to
the $R_{xx}$ or $\Delta R_{xx}$ oscillations, as shown in Fig.
1(d)[bottom]. Here, we survey these experimentally observed phase
relations and related crossovers in the high mobility 2DES, and
then focus on the "Type-3" case, which also brings with it,
remarkably, IQHE.

Simultaneous lock-in based electrical measurements of the diagonal
resistance and the Hall effect were carried out at $T > 0.45 K$,
with matched lock-in time constants, and sufficiently slow
$B$-field sweep rates. The $B$-field was calibrated by ESR of
DPPH.\cite{6} We are able to rule out spurious contributions in
the phase shift, between $\Delta R_{xx}$ and $\Delta R_{xy}$,
originating from the B-sweep rate, the data acquisition rate,
lock-in integration time, and other typical experimental
variables.  As usual, the observed magnetoresistance oscillations
became weaker at higher $T$, and few oscillations were evident for
$\nu > 20$ at $T
> 1.7 K$. Thus, this study focused upon $0.45 < T < 1.7 K$, where
$T$-induced changes in the phase relations were not discerned. The
observed phase relations also did not show an obvious dependence
on the sample geometry. We note, parenthetically, that the phase
relation between the oscillatory Hall- and diagonal- resistances
could often be identified by the trained eye. Thus, background
subtraction helps mainly to bring out the Hall oscillations from
$R_{xy}$, and make possible $\Delta R_{xx}$, $\Delta R_{xy}$
overlays, for phase comparison. Typically, $R_{xx}^{back}$
followed either the midpoints (e.g. Fig. 1(a)) or the minima (e.g.
Fig. 2(d)) of the $R_{xx}$ oscillations. For $R_{xy}^{back}$,
since $|R_{xy}^{back}|
>> |\Delta R_{xy}|$, we used a two pass procedure where the first
pass identified $\approx 99\%$ of $R_{xy}^{back}$ through a
linear-fit of $R_{xy}$, and a spline fit in the second pass then
accounted for the $\approx 1\%$ residual term. Here, at the second
pass, $R_{xy}^{back}$ was optimized to make possible $\Delta
R_{xx}$, $\Delta R_{xy}$ overlays. Finally, although $\mu$ has
been provided, $\mu$ alone seems to be insufficient for
classifying the observed phenomena in high-$\mu$ specimens. Here,
the high mobility condition was realized in GaAs/AlGaAs by brief
illumination with a red LED.

\begin{figure}[h]
%h=here, t=top, b=bottom, p=separate figure page
\begin{center}
\leavevmode \epsfxsize=2.5in
 \epsfbox {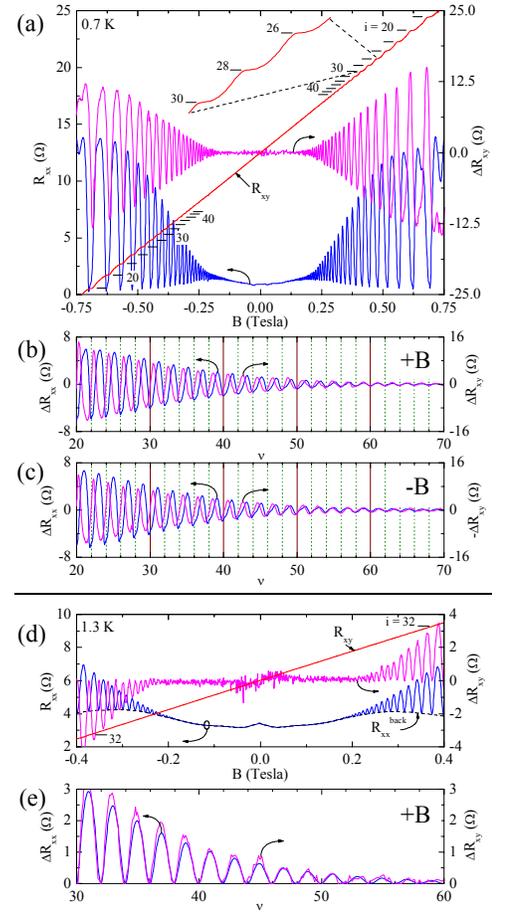}
\end{center}
\caption{(a) $R_{xx}$, $R_{xy}$, and the oscillatory Hall
resistance, $\Delta R_{xy}$, are shown over low magnetic fields,
$B$ for a high mobility square shape GaAs/AlGaAs specimen. Here,
$\Delta R_{xy}(-B) = -\Delta R_{xy}(+B)$. (b) The oscillatory
diagonal resistance ($\Delta R_{xx}$) and $\Delta R_{xy}$ have
been plotted vs. $\nu$ to compare their relative phases for $+B$.
For $20 \leq \nu < 46$, $\Delta R_{xx}$ and $\Delta R_{xy}$ are
approximately orthogonal as in Fig. 1(c). For $56 \leq \nu \leq
70$, $\Delta R_{xx}$ and $\Delta R_{xy}$ are approximately
in-phase, unlike at $\nu < 46$. (c) As above for $-B$. Note that
the right ordinates in Fig. 2(b) and Fig. 2(c) show $+\Delta
R_{xy}$ and $-\Delta R_{xy}$, respectively, in order to account
for the antisymmetry in $\Delta R_{xy}$ under $B$-reversal. (d)
$R_{xx}$, $R_{xy}$, and $\Delta R_{xy}$ have been shown for a Hall
bar specimen with $n = 2.9 \times 10^{11} cm^{-2}$ and $\mu = 6
\times 10^{6} cm^{2}/V s$, which exhibits in-phase $R_{xx}$ and
$\Delta R_{xy}$ oscillations. Note the absence of discernable Hall
plateaus in $R_{xy}$. (e) Here, the amplitudes of $\Delta R_{xx}$
and $\Delta R_{xy}$ show similar $\nu$ -variation. }
\label{mani02fig}
\end{figure}

For a high mobility square shaped specimen with $n = 3 \times
10^{11} cm^{-2}$ and $\mu = 1.1 \times 10^{7} cm^{2}/V s$ that
shows IQHE up to $i \approx 40$, Fig. 2(a) illustrates $R_{xx}$,
$R_{xy}$, and $\Delta R_{xy}$ for both $B$-directions, using the
convention $R_{xy} > 0$ for $B > 0$. Figures 2(b) and (c) confirm
similar behavior for both B-directions once the anti-symmetry in
$\Delta R_{xy}$ under $B$-reversal is taken into account. Fig.
2(b) indicates that from $20 \leq \nu < 46$, $\Delta R_{xy}$
oscillations are approximately orthogonal to the $\Delta R_{xx}$
oscillations, as in Fig. 1(c). This feature, and the manifestation
of Hall plateaus in Fig. 2(a), confirms that the IQHE observed
here is the canonical effect. A remarkable and interesting feature
in Fig. 2(b) is that, following a smooth crossover, $\Delta
R_{xx}$  and $\Delta R_{xy}$ become in-phase, i.e., "Type-2", for
$\nu \geq $ 56. That is, with increasing $B$ (or decreasing
$\nu$), the system exhibits a Type-2 $\rightarrow$ Type-1
transformation.

\begin{figure}[t]
%h=here, t=top, b=bottom, p=separate figure page
\begin{center}
\leavevmode \epsfxsize=2.5in
 \epsfbox {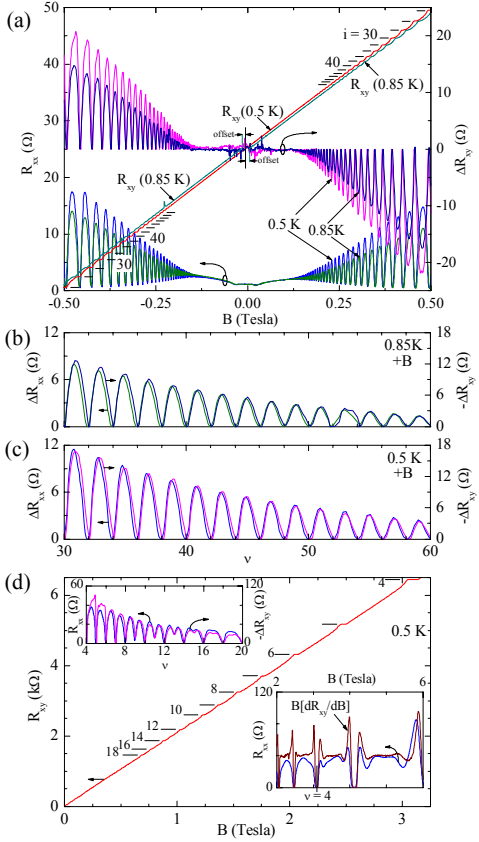}
\end{center}
\caption{(a) Data for a square shape GaAs/AlGaAs specimen, where
the magnitude of $R_{xy}$ is suppressed at the $R_{xx}$
oscillation maxima. Here, the $+B$ and the $-B$ portions of the
$R_{xy} (0.85 K)$ curve have been offset in opposite directions
along the abscissa with respect to the $R_{xy} (0.5 K)$ curve, for
the sake of presentation. Note the well-developed plateaus in
$R_{xy}$. (b) At $T= 0.85 K$, a plot of $\Delta R_{xx}$ and
$-\Delta R_{xy}$ confirms a Type-3 phase relation. (c) Same as (b)
but at $T= 0.5 K$. (d) The main panel shows the Hall resistance
vs. $B$, with relatively narrow Hall plateaus, down to around
filling factor $\nu = 4$. Left Inset: As in Fig. 3(b) and 3(c)
above, $-\Delta R_{xy}$ follows $R_{xx}$, down to nearly $\nu =
5$. Right Inset: For $\nu \leq 4$, a better correspondence
developed between $R_{xx}$ and $B[dR_{xy}/dB]$. }
\label{mani03fig}\end{figure}

Figs. 2(d) and 2(e) provide further evidence for in-phase Type-2
oscillations in a Hall bar. Here, the Hall oscillations tend to
enhance the magnitude of $R_{xy}$ at the $R_{xx}$ SdH maxima
("Type-2"), even as Hall plateaus are imperceptible in the
$R_{xy}$ curve. Yet, from Fig. 2(d), it is clear that $\Delta
R_{xy}$ is a Hall effect component, and not a misalignment offset
admixture of $R_{xx}$ into $R_{xy}$, since $\Delta R_{xy}$ is
antisymmetric under $B$-reversal.
\begin{figure}[t]
%h=here, t=top, b=bottom, p=separate figure page
\begin{center}
\leavevmode \epsfxsize=2.5in \epsfbox {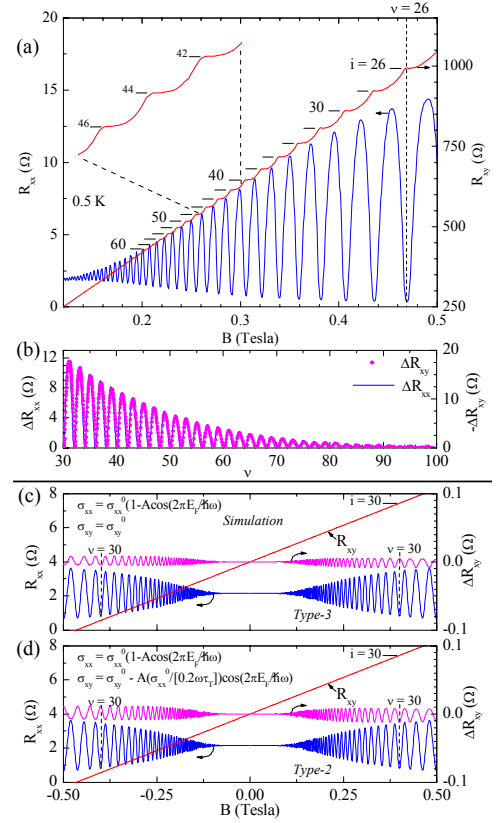}
\end{center}
\caption {(a) $R_{xx}$ and $R_{xy}$ in a high mobility GaAs/AlGaAs
specimen that exhibits deep $R_{xx}$ minima and even-integral Hall
plateaus at $R_{xy} = h/ie^{2}$. (b) Here, $-\Delta R_{xy}$
follows $\Delta R_{xx}$, indicative of a $\pi$ phaseshift and a
"Type-3" relationship, unlike the typical quantum Hall situation,
see Fig. 1(c). (c) Simulations suggest that oscillatory scattering
corrections to the diagonal conductivity alone can produce Type-3
oscillations, via the tensor relations for the resistivities. (d)
Simulations of a semiempirical model that includes both a
scattering contribution in $\sigma_{xy}$, and a reduction in the
relaxation time with respect to the transport lifetime, indicate
the possibility also of Type-2 oscillations, with $|\Delta R_{xy}|
<< |\Delta R_{xx}|$. A comparison of Fig. 4(c) and 4(d) helps to
convey the $\tau$-induced Type-3 to Type-2 transformation.}
\label{mani05fig}
\end{figure}

Figure 3 illustrates the third ("Type-3") phase relation in a high
mobility square shape 2DES with $n = 2.9 \times 10^{11} cm^{2}$
and $\mu = 1 \times 10^{7} cm^{2}/V-s$. Although $\mu$ for this
specimen is similar to the one examined in Fig. 2(a)-(c), the
experimental results do look different. Figure 3(a) exhibits data
taken at $T = 0.85 K$ and $T = 0.5 K$. Fig. 3(a) shows that the
main effect of changing $T$ is to modify the amplitude of the
$\Delta R_{xx}$ and $\Delta R_{xy}$ oscillations, so that
oscillatory effects persist to a lower $B$ at the lower $T$. The
data of Fig. 3(a) also show that $\Delta R_{xy}$ tends to reduce
the magnitude of $R_{xy}$ over the B-intervals corresponding to
the $R_{xx}$ peaks, as in Fig. 1d(bottom), the Type-3 case.
Meanwhile, quantum Hall plateaus are easily perceptible in
$R_{xy}$, see Fig 3(a). Figures 3(b) and 3(c) demonstrate that for
$+B$, for example, $\Delta R_{xx}$ and $-\Delta R_{xy}$ show the
same lineshape for $30 \le \nu \le 60$, and the phase relation
does not change with $T$. Indeed, this correlation held true down
to nearly $\nu = 5$, see left inset of Fig. 3(d), as narrow IQHE
plateaus were manifested in $R_{xy}$, see Fig. 3(d). For $\nu \le
4$, however, $R_{xx}$ correlated better with $B[dR_{xy}/dB]$, see
right inset of Fig. 3(d), than with $- \Delta R_{xy}$, which
suggested that the resistivity/resistance rule\cite{10} comes into
play at especially low-$\nu$ here, as the system undergoes a
Type-3 $\rightarrow$ Type-1 transformation, with decreasing $\nu$.

An expanded data plot of Type-3 transport is provided in Fig.
4(a). This plot shows plateaus in $R_{xy}$ and deep minima in
$R_{xx}$ to very low-B, as quantum Hall plateaus in $R_{xy}$
follow $R_{xy} = h/ie^{2}$, to an experimental uncertainty of
$\approx 1$ percent. Although the IQHE data of Fig. 4(a) again
appear normal at first sight, the remarkable difference becomes
apparent when $\Delta R_{xx}$ and -$\Delta R_{xy}$ are plotted vs.
$\nu$, as in Fig. 4(b). Here, we find once again a phase-shift of
"$\pi$" ("Type-3") between  $\Delta R_{xx}$ and $\Delta R_{xy}$
(Fig. 4(b)), as in Fig. 3(b) and (c), that is distinct from the
canonical ("Type-1") phase relationship exhibited in Fig. 1(c). In
Fig. 4(a), the reported phase relation can even be discerned by
the trained eye.

It is possible to extract some insight from the phase
relationships observed here between $R_{xx}$ (or $\Delta R_{xx}$)
and $\Delta R_{xy}$. The Type-1 orthogonal phase relation of Fig.
1(c) can be viewed as a restatement of the empirical
resistivity/resistance rule, since the data of Fig. 1(a) yield
both Fig. 1(b) and Fig. 1(c). Theory suggests that this rule might
follow when $R_{xx}$ is only weakly dependent on the local
diagonal resistivity $\rho_{xx}$ and approximately proportional to
the magnitude of fluctuations in the off-diagonal resistivity
$\rho_{xy}$, when $\rho_{xx}$ and $\rho_{xy}$ are functions of the
position.\cite{14} Thus, specimens exhibiting the resistivity rule
(and Type-1 oscillations) seem likely to reflect density
fluctuations/inhomogeneities.\cite{14}

For Type-2 and Type-3 oscillations, note that the specimens of
Figs. 2 - 4 satisfy $\omega \tau_{T} > 1$, with $\omega$ the
cyclotron frequency, and $\tau_{T}$ the transport lifetime, at $B
> 0.001$ (or $0.002$) $T$. Yet, one might semi-empirically
introduce oscillations into the diagonal conductivity,
$\sigma_{xx}$, as $\sigma_{xx} = \sigma_{xx}^{0}(1 - Acos(2\pi
E_{F}/\hbar\omega ))$.\cite{15,16,17} Here, the minus sign ensures
the proper phase, while $\sigma_{xx}^{0} = \sigma_{0}/(1+(\omega
\tau_{T})^{2})$, $\sigma_{0}$ is the $dc$ conductivity, $E_{F}$ is
the Fermi energy, and $A = 4c[(\omega \tau_{T})^{2}/(1+(\omega
\tau_{T})^{2})][X/sinh(X)]exp(- \pi/\omega \tau_{S})$, where $X =
[2 \pi^{2}k_{B}T/\hbar \omega]$, $\tau_{S}$ is single particle
lifetime, and $c$ is of order unity.\cite{15,17,18} Simulations
with $\sigma_{xx}$ as given above, and $\sigma_{xy} =
\sigma_{xy}^{0} =-(\omega \tau_{T})\sigma_{xx}^{0}$, indicate
oscillations in both $R_{xx}$ and $R_{xy}$ via $\rho_{xx} =
\sigma_{xx} /(\sigma_{xx}^{2}+ \sigma_{xy}^{2})$ and $\rho_{xy} =
\sigma_{xy} /(\sigma_{xx}^{2}+ \sigma_{xy}^{2})$, and a Type-3
phase relationship, see Fig. 4(c), with $|\Delta R_{xy}| <<
|\Delta R_{xx}|$. That is, an oscillatory $\sigma_{xx}$ can also
lead to $R_{xy}$ oscillations.

As a next step, one might introduce an oscillatory $\sigma_{xy} =
\sigma_{xy}^{0}(1 + Gcos(2 \pi E_{F}/ \hbar \omega))$, where $G =
2c[(1+3( \omega \tau_{T})^{2})/((\omega \tau_{T})^{2}(1+(\omega
\tau_{T})^{2}))] [X/sinh(X)]exp(-\pi/\omega
\tau_{S})$.\cite{16,17} Upon inverting the tensor including
oscillatory $\sigma_{xy}$ and $\sigma_{xx}$, Type-3 oscillations
were still obtained, as in Fig. 4(c).

Finally, the strong $B$-field $\sigma_{xy}$ follows $\sigma_{xy} =
\sigma_{xx}/\omega \tau - ne/B$ in the self-consistent Born
approximation for short range scattering potentials, when $\tau$
is the relaxation time in the $B$-field.\cite{16} Although, the
dominant scattering mechanism is long-ranged in GaAs/AlGaAs
devices, we set $\sigma_{xy} = (\sigma_{xx}^{0}/\omega \tau -
ne/B) - A(\sigma_{xx}^{0}/\omega \tau)cos(2\pi E_{F}/\hbar
\omega)$. When $\tau = \tau_{T}$, this approach again yielded
Type-3 phase relations, as in the discussion above. Next, we
examined the case $\tau < \tau_{T}$, in order to account for the
possibility that $\tau$ in a $B$-field may possibly come to
reflect $\tau_{S}$, which typically satisfies $\tau_{S} <
\tau_{T}$ for small angle scattering by long-range
scattering-potentials.\cite{18} Remarkably, a reduction in $\tau$,
which corresponds to changing the nature of the potential
landscape, converted Type-3 (phase-shift by $\pi$) to Type-2
(in-phase) oscillations, see Fig. 4(c) and 4(d).

If density fluctuations at large length scales produce Type-1
characteristics,\cite{14} and Type-2 oscillations require a
difference between $\tau_{T}$ and $\tau$ as suggested above, then
the observation of Type-1 and Type-2 oscillations in the same
measurement (Fig. 2(b) and (c)) is consistent because long
length-scale potential fluctuations can produce both modest
density variations and a difference between $\tau_{T}$ and
$\tau_{S}$ (or $\tau$).\cite{18} Perhaps, with increasing $B$,
there is a crossover from "Type-2" to "Type-1" before $R_{xy}$
plateaus become manifested, and thus, IQHE is not indicated in the
Type-2 regime, see Fig. 2 (a) and (d).

Specimens exhibiting Type-3 oscillations and associated IQHE
suggest better homogeneity in $n$, which is confirmed by
oscillations to extremely low-$B$ (see Fig. 4(a) and (b)), to
nearly $\nu = 100$ at $T = 0.5$ K. The relatively narrow plateaus
at high-$B$ (see Fig. 3(d)) hint at a reduced role for
disorder-induced localization.\cite{2} In this case, perhaps there
are other new mechanisms contributing to the large $|\Delta
R_{xy}|$, and plateau formation, in the high-$\mu$ system. It
could be that new physics serves to create/maintain/enhance the
mobility gap or suppress backscattering in the higher Landau level
here, which assists in the realization of the "Type-3"
characteristics and IQHE to very low $B$. As theory and experiment
have already suggested such possibilities in the higher Landau
levels,\cite{4,5,9,10}, such ideas seem to merit
consideration.\cite{19}

\end{document}